\documentclass[preprint,prd]{revtex4-1}
\usepackage{color}
\usepackage{graphicx}
\usepackage{epsfig}
\usepackage{subfig}
\usepackage{lipsum}
\usepackage{ctable}
\usepackage{hyperref}
\usepackage{cleveref}
\crefformat{footnote}{#2\footnotemark[#1]#3}

\usepackage{graphicx}
\usepackage{dcolumn}
\usepackage{bm}

\begin{document}

\title{ Radiation Fluid  Stars in    the Non-minimally Coupled $Y(R)F^2$ Gravity}

\author{ \"{O}zcan SERT}
\email{osert@pau.edu.tr}
 \affiliation{Department of Mathematics, Faculty of Arts and Sciences,
Pamukkale
University,  20070   Denizli, T\"{u}rkiye}


\date{\today}

\begin{abstract}

 \noindent
 
 We propose a non-minimally coupled gravity model  in  $ Y(R)F^2 $  form to describe the radiation fluid stars which have  the radiative equation of state between the energy density $\rho$ and the pressure $p$ as $\rho=3p\;$. 
  Here  
 $F^2$ is the Maxwell invariant and  $Y(R)$  is a function of the Ricci scalar  $R$.
We give the gravitational and electromagnetic field equations  in      differential form notation taking the infinitesimal variations  of the  model. We look for  electrically charged star solutions to the field equations under a constraint  which eliminating complexity of the higher order terms in the field equations. We determine the non-minimally coupled function $Y(R)$  and  the corresponding model which  admits   new  exact   solutions in   the interior   of star and Reissner-Nordstrom solution at the exterior region. 
 Using vanishing pressure  condition at the boundary together with the  continuity conditions  of the metric functions and the  electric charge, we find  the mass-radius ratio, charge-radius ratio and gravitational surface redshift depending on   the parameter of the model for the radiation fluid  star.
 We derive 
 general restrictions   for the   ratios and redshift of  the charged compact stars. We obtain
 a slightly smaller upper mass-radius ratio  limit  than the Buchdahl bound $4/9\;$ and a  smaller upper redshift limit  than the bound of the standard General Relativistic stars.



\end{abstract}

\pacs{Valid PACS appear here}
\maketitle


\def\ba{\begin{eqnarray}}
\def\ea{\end{eqnarray}}
\def\w{\wedge}



\section{Introduction}

\noindent

The radiation fluid stars have  crucial importance in astrophysics. 
They  
can describe the core of neutron stars which is   a collection of  cold degenerate (non-interacting)
fermions  \cite{oppen,misner,meltzer,chandrasekhar}   and   self-gravitating photon stars \cite{chandrasekhar,sorkin,schmidt,chavanis1,chavanis2}.
Such radiative stars which called “Radiation Pressure
Supported Stars” (RPSS)  can be  possible even in   Newtonian Gravity\cite{mitra} and their  relativistic extension which called  
“Relativistic Radiation Pressure Supported Stars” (RRPSS)  \cite{glend}    can describe the  
gravitational collapse of massive matter clouds to very high density fluid. There are also some
investigations  
related  with gedanken experiments
such as black hole formation and evaporation with self-gravitating  gas 
confined by a  spherical symmetric box. These investigations   \cite{penrose,wald}  can lead to new insights to the nature of Quantum Gravity.

The  radiation fluid stars have
the radiative equation of state    with $\rho= 3p $ which is the high density limit of the general isothermal
spheres satisfying  the linear barotropic equation of state $\rho= kp $ with constant $k$.
 The entropy and thermodynamic stability of    self-gravitating charge-less radiation fluid  stars 
were firstly calculated in \cite{sorkin} using
  Einstein   equations. This work was   extended to 
 the investigation of structure,  stability and  thermodynamic parameters of the isothermal  spheres  involving photon  stars and the core of neutron stars   \cite{chavanis1,chavanis2}.  
Also, the numerical  study of  such a charge-less radiative star which consists of a photon gas conglomerations    can   be  found in \cite{schmidt}.
Some interesting interior solutions of the General Relativistic field
 equations  in isotropic coordinates with the linear barotropic equation of state   were presented by Mak and Harko \cite{mak-harko-2005,mak-2013,harko2016}  for dense  astrophysical objects without charge.

  Furthermore,
  a spherically symmetric fluid  sphere  which contains a constant surface charge can be more stable than the charge-less case \cite{Stettner}.
 The gravitational collapse of a spherically
 symmetric     star  
 may  be prevented by   charge \cite{Krasinski}, since the repulsive electric force contributes  to counterpoise the gravitational attraction \cite{Sharma}. 
 It is interesting to note that   
 interior of a  strange quark star  can  be described by a  charged solution 
admitting  a one-parameter group of conformal motions \cite{mak-harko-2004}  for the equation of state  $\rho = 3p + 4B $ which  is known as  the MIT bag model. 
The physical  properties and structure of the radiation fluid stars in the model with hybrid metric-Palatini gravity \cite{danila-harko}  and  Eddington-inspired
 Born-Infeld (EIBI) gravity \cite{harko-lobo-2013}      were obtained   numerically.
 It is a challenging problem to find exact interior solutions   of the the charged radiation fluid stars, since the trace of the gravitational  field equations gives zero Ricci scalar    for    the radiative equation of state $\rho=3p$. Therefore  it is important to find a modified  gravity  model which  can describe  the radiation fluid stars analytically.


In this study  we propose a non-minimally coupled modified gravity model in $ Y(R)F^2 $-form  in order to  find exact solutions to the radiation fluid stars. Here  
$F^2$ is the Maxwell invariant and  $Y(R)$  is a function of the Ricci scalar  $R$. We   will determine the non-minimal function from   physically applicable solutions of   the field equations and boundary conditions. Such  a coupling in $RF^2$ form 
was first introduced by Prasanna \cite{Prasanna}  to understand  the 
intricate nature between all energy forms, electromagnetic fields  and curvature. Later, a class of such  couplings  was investigated  to gain more insight on charge conservation and curvature \cite{Horndeski}.
These non-minimal terms can be  obtained
from  dimensional  reduction of   a five dimensional  Gauss-Bonnet  gravity action \cite{Mueller} and $R^2$-type   action \cite{Buchdahl,Dereli90}.  The  calculation of QED  photon propagation  in a curved background metric \cite{Drummond} leads to these terms. 
A generalization of the non-minimal model to $R^nF^2$-type  couplings \cite{Turner,Campa,Kunze,Mazzi,dereli4,bamba1}
may  explain the generation of seed magnetic fields during the inflation and 
 the origin of large-scale magnetic fields 
 in the universe \cite{Turner,Mazzi,Campa}. 
Another generalization  of the non-minimal $RF^2$ model
 to    non-Riemannian
 space-times \cite{Baykal} can give   more insights to the torsion and electromagnetic fields.
 Then it is possible to consider the more general couplings with  any function of the Ricci scalar   and the electromagnetic fields
such as  $Y(R) F^2$-form. These non-minimal models in $Y(R) F^2$-form    have very interesting solutions such as regular black hole solutions to avoid singularity \cite{Sert16regular}, spherically symmetric static  solutions to explain  the rotation curves of galaxies \cite{Sert12Plus,dereli3,dereli4,Sert13MPLA}, cosmological solutions 
 to explain cosmic acceleration    of the universe \cite{Sert-Adak12,AADS,bamba1,bamba2} and pp-wave solutions \cite{dereli2}.
 
 In order  to investigate  a  charged  astrophysical  phenomena  one can consider Einstein-Maxwell theory which is a minimal coupling between gravitational and electromagnetic fields. But  when the astrophysical  phenomena has high density, pressure and  charge  such as neutron stars and quark stars,
 new interaction types between gravitational and electromagnetic fields may be  appeared. 
 Then the non-minimally coupled  $Y(R)F^2$ Gravity can be ascribed to such new interactions
  and  we can apply    the  theory 
to    the charged compact stellar system.  In this study we focus on exact solutions of the radiation fluid stars for the non-minimally coupled model,
inspired by the solution in \cite{mak-harko-2004}. We construct the non-minimal coupling function  $Y(R)$  with  the parameter  $\alpha$ and   the corresponding model. We give interior and exterior solutions of the model. Similarly to \cite{mak-harko-2004}, our interior  solutions turn out to the solution given by Misner and Zapolsky \cite{misner} with $b=0$ and $Q=0$, 
describing  ultra high density neutron star  or the relativistic Fermi gas.   We determine the total mass, charge  and surface gravitational  redshift of the stars depending  on the boundary radius  $r_b$  and the parameter $\alpha$ using the matching conditions.  
We give 
 the general restrictions   for the   ratios and redshift of  the charged compact stars and compare them with the bound given  in \cite{Mak1} and 
  the Buchdahl bound \cite{Buchdahl2}.

The organization of the present work   is  as follows:  
The general  action in $Y(R)F^2$ form  and the corresponding field  equations are given in Section II to  describe a charged compact star.   The  spherically symmetric, static exact solutions under   conformal symmetry  and    structure of the non-minimal function $Y(R)$     are obtained in Section III.  Using the continuity  and  boundary conditions,  the gravitational mass, total charge and redshift of the star are derived  in Section IV. The conclusions are given in the last Section.

\section{ The  Model with $Y(R)F^2 $-Type Coupling for a Compact Star} \label{model}

The recent astronomical observations about the problems such as  dark matter \cite{Overduin,Baer} and dark energy \cite{Riess,Perlmutter,Knop,Amanullah,Weinberg,Schwarz}  strongly support that  Einstein's theory of gravity needs to modification  at large scales. Therefore
 Einstein-Maxwell theory also may be modified \cite{AADS,dereli3,Sert12Plus,Sert13MPLA,bamba1,bamba2,Sert-Adak12,dereli4,Turner,Mazzi,Campa} 
 to explain these observations. 
Furthermore, since such  astrophysical  phenomenons  as neutron stars or quark stars have high energy density, pressure and  electromagnetic fields,
new interaction types between gravitational and electromagnetic fields in  $Y(R)F^2$ form can be appeared.  When the extreme conditions are  removed, this model turns out to the minimal Einstein-Maxwell theory.
  We  write  the following   action    to describe    the interior of a charged compact  star  by adding the matter part $L_{mat}$ and the source term $A\wedge J$ to   the  $Y(R)F^2 $-type non-minimally coupled model in \cite{dereli3,dereli4,Sert13MPLA, Sert12Plus,Sert16regular,Sert-Adak12,AADS,bamba2},
\begin{equation}\label{action}
        I = \int_M{\{   \frac{1}{2\kappa^2} R*1 -Y(R) F\w *F + 2A\wedge J + L_{mat} + \lambda_a\wedge T^a  \} }
        \nonumber
\end{equation}
 depending on  the fundamental variables such that   the co-frame 1-form $\{e^a\}$,
  the  connection 1-form   ${\{\omega^a}_b\}$ and   the homogeneous electromagnetic
  field 2-form $F $.  We  derive  $F$ from the electromagnetic  potential 1-form $A$ such as $F=dA$.
   We  constrain the model to the case with zero torsion connection  by $\lambda_a$
  Lagrange multiplier 2-form.
   Then the variation of $\lambda_a$ leads to the   Levi-Civita connection which 
 can be found from $T^a =  de^a + \omega^{a}_{\;\;b} \w e^b=0$.  
In  action (\ref{action}),   $J$  is the  
electric current density 3-form for  the  source fluid inside  the star, $Y(R)$ is any function of    the curvature  scalar $R$. The  scalar   can be derived   from the curvature tensor 2-forms   $ R^{a}_{\;\;b} = d\omega^{a}_{\;\;b} + \omega^{a}_{\;\;c} \w \omega^{c}_{\;\;b} $
  via the interior product $ \iota_{a}$ \  such as $    \iota_{b} \iota_{a} R^{ab}= R $.  We denote the space-time metric by $g = \eta_{ab} e^a \otimes 
  e^b$  which has  the signature  $(-+++)$. Then we set
  the volume element  with
  $*1 = e^0 \w e^1 \w  e^2 \w e^3 $  on the four dimensional manifold.

  For a charged isotropic perfect fluid, Electromagnetic and  Gravitational   field equations of the non-minimal  model are found from    infinitesimal variations   of the action (\ref{action})
  \begin{eqnarray}
   d(*Y F) &=& J \; ,   \ \label{maxwell1}
   \\   \   \   dF&=& 0 \;, \ 
  \\ \label{gfe}
  - \frac{1}{2 \kappa^2}  R_{bc}
 \w *e^{abc}  & = &   Y  (\iota^a F \w *F - F \w \iota^a *F)   +  Y_R F_{mn} F^{mn}*R^a 
 \nonumber
 \\
 && +   D [ \iota_b\;d(Y_R F_{mn} F^{mn} )]\wedge *e^{ab} + \  (\rho +p )u^a*u + p*e^a
 \  ,
 \end{eqnarray}
 where  $Y_R = \frac{dY}{dR}$ and $u= u_a e^a $  is the velocity 1-form   associated with an  inertial time-like observer,  $u_au^a =-1 $. 
 The modified Maxwell equation (\ref{maxwell1}) can be written as 
 \begin{eqnarray}\label{d*G}
 d*\mathcal{G} = J
 \end{eqnarray}
 where $\mathcal{G}=YF$ is the excitation 2-form in
 the interior medium of the star.
  More detailed analysis on this subject  can be found in \cite{dereli2,Dereli5,Dereli6}. 
   Following Ref. \cite{AADS} we write the gravitational field equations (\ref{gfe}) as follows:
   \begin{eqnarray}\label{Ga}
 G^a  =  \kappa^2 \tau^a_{N}  +  \kappa^2 \tau^a_{mat},
 \end{eqnarray}
 where $G^a  $ is Einstein tensor  
 $
 G^a =
 - \frac{1}{2 }  R_{bc}
 \w *e^{abc} \;,
 $
$\tau^a_{N} $    and  $\tau^a_{mat}$ are two separate effective    energy momentum tensors, namely,   the energy momentum tensor of the   non-minimally coupled term    introduced in \cite{AADS,Sert16regular} and   energy momentum tensor of matter, respectively
  \begin{eqnarray}
 \tau^a_{N}  &=&  Y  (\iota^a F \w *F - F \w \iota^a *F)   + Y_R F_{mn} F^{mn}*R^a 
 \nonumber\label{tauan}
 \\ && +  D [ \iota_b\;d(Y_R F_{mn} F^{mn} )]\wedge *e^{ab} \;,
 \\
 \tau^a_{mat} &=& \frac{\delta L_{mat}}{\delta e_a }  =  (\rho +p )u^a*u + p*e^a\;. \label{tauam}
 \end{eqnarray}
 
 We take the exterior covariant derivative of the  modified gravitational equation (\ref{Ga}) in order to show that the conservation of the total energy-momentum tensor   $\tau^a =  \tau^a_{N}  +    \tau^a_{mat} $ 
 \begin{eqnarray}\label{DGa}
 DG^a = D\tau^a \;.
 \end{eqnarray}
 
 The left hand side of equation (\ref{DGa}) is identically zero $DG^a = 0$. The right hand side of equation (\ref{DGa})
   is calculated term by term as follows:
\begin{eqnarray}
 D[ Y  (\iota^a F \w *F - F \w \iota^a *F)]  &=&  2 J\wedge F^a -\frac{1}{2} F_{mn}F^{mn}dY \wedge *e^a\\
 D(Y_RF_{mn}F^{mn} *R^a) &=& d(Y_RF_{mn}F^{mn}) \wedge *R^a  + \frac{1}{2}F_{mn}F^{mn} Y_RdR\wedge *e^a\\
D\left[ D[ \iota_b d(Y_R F_{mn} F^{mn} )]\wedge *e^{ab} \right] &=& D^2 [ \iota_b d(Y_R F_{mn} F^{mn} )]\wedge *e^{ab} \\
&=& {R_b}^c\wedge \imath_c\;d(Y_R F_{mn} F^{mn} )  *e^{ab} \\
&=& -d(Y_R F_{mn} F^{mn} )\wedge *R^a\;.
 \end{eqnarray}
If we substitute all the expressions 
in (\ref{DGa})  we find
    \begin{eqnarray}
   0 = D\tau^a = 2J\wedge F^a +  D\tau^a_{mat}
    \end{eqnarray}
    which leads to
    \begin{eqnarray}\label{DtauaJ}
  D\tau^a_{mat} =-2J\wedge F^a
    \end{eqnarray}
which is similar to   the minimally coupled Einstein-Maxwell theory, but where $J = d(*YF)$ from (\ref{maxwell1}).  Then in  this case without source $J=0$, the conservation of  energy-momentum tensor becomes $ D\tau^a = 0 =  D\tau^a_{mat}\;.$
   
The isotropic matter   has the following energy density and pressure    $  \rho =  \tau_{mat}^{0,0} $, \ 
  $  p= \tau_{mat}^{1,1} = \tau_{mat}^{ 2,2} = \tau_{mat}^{3,3} $
  as the diagonal components of the matter  energy momentum tensor $\tau^a_{mat}$ in  the interior of the star.  
 In order to  get over  higher order derivatives and the complexity of  the last term in  (\ref{tauan})  
we take  
the following constraint 
\begin{eqnarray}\label{cond0}
Y_R F_{mn} F^{mn} = \frac{K}{ \kappa^2}  \label{YRFk}
\end{eqnarray}
where $K $  is a non-zero  constant. If one take $K$ is zero, then the non-minimal function $Y$ becomes a constant and this  is not different from the well known  minimally coupled   Einstein-Maxwell theory. The constraint (\ref{cond0}) has the following  futures: First of all,  this constraint (\ref{cond0})   is  not an independent equation from the field equations, since   the exterior covariant derivative  of the  gravitational field equations  under the condition gives  the constraint again in addition to  the conservation equation.
 Secondly, the field equations (\ref{maxwell1}-\ref{gfe}) under  the  condition  (\ref{cond0}) with $K=-1$, 
 can be interpreted \cite{AADS} as the field equations of
the trace-free Einstein gravity \cite{Einstein,Weinberg2}  or unimodular gravity  \cite{Unruh,Ellis}   which  coupled to electromagnetic 
energy-momentum tensor  with the non-minimal function $Y(R)$,
which are viable for astrophysical  and  cosmological applications. 
Thirdly, the constraint allows us  to find  the   other physically interesting solutions of the non-minimal model \cite{AADS,Sert16regular,Sert13MPLA,dereli3,dereli4,bamba1,bamba2}.     Fourthly,  when we take the trace of the gravitational field equation (\ref{gfe})  as done  in Ref. \cite{AADS},  we obtain 
  \begin{eqnarray}
  \frac{K+1}{\kappa^2} R*1 = (\rho  - 3p)*1 \;. \label{trace}
  \end{eqnarray}
We can consider two cases  satisfying (\ref{trace})  for the non-minimal $Y(R)F^2$   coupled model:
 \begin{enumerate}
 		\item  $K=-1$ which leads to  the equation of state $\rho = 3 p$  for the radiation fluid stars.
 	 \item	$K\neq - 1  $
 	with  the equation $R = \frac{\kappa^2 ( \rho- 3p)}{K+1}$.
 	\end{enumerate}
 Then we set  $K= -1$ in (\ref{cond0}) and (\ref{trace}), since  we concentrate on the radiation fluid star for   the non-minimal model. Therefore we see that the trace of the  gravitational field equations  does not give a new independent equation as another feature of the  condition (\ref{cond0}) with $K=-1$.
One may see Ref. \cite{AADS} for a detailed discussion on the physical properties and features of $\tau_{N}^{a}$ for the case $K=-1$. We  leave the  second   case with $K\neq - 1 $, $\rho \neq 3p$   for next studies.  We also note that the non-minimally coupled $Y(R)F^2$ model does not give  any new solution for the MIT bag model $\rho -3p = 4B$   with $B \neq 0$, since  the curvature scalar $R$ becomes a constant in (\ref{trace}) therewith   $Y(R)$ must be constant. Thus this case is not a new model but the minimal Einstein-Maxwell case.

 \section{Static, Spherically Symmetric Charged  Solutions}

 \noindent We seek  solutions to the model with $Y(R)F^2$-type coupling  describing  a  radiative compact star for the    following   most general (1+3)-dimensional  spherically symmetric,  static metric 
 \begin{equation}\label{metric}
 ds^2 = -f^2(r)dt^2  +  g^2(r)dr^2 + r^2d\theta^2 +r^2\sin(\theta)^2 d \phi^2
 \end{equation}
 and    the following  electromagnetic tensor 2-form with the  electric field component $E(r)$   
 \begin{eqnarray}\label{electromagnetic1}
 F     = E(r) e^1\wedge e^0 .
 \end{eqnarray}
We take the electric current density  as a source of the field which has only   the electric charge density  component $\rho_e(r)$
\begin{eqnarray}
J = {\rho} _e(r) e^{1}\wedge e^{2} \wedge e^{3} = {\rho} _e g r^2 \sin\theta dr \wedge d\theta \wedge d\phi\;.
\end{eqnarray}
Using  Stokes theorem, integral form of the Maxwell  equation (\ref{maxwell1})  can be written as
\begin{eqnarray}\label{stokes}
\int_{V} d*YF = \int_{\partial V} *YF = \int_V J =4\pi q(r)
\end{eqnarray}
  over  the sphere which has  the volume  $V$ and  the boundary  $\partial V$.
  When we take the integral, we find  the  charge  inside  the volume with the radius $r$
  \begin{eqnarray}
  YEr^2  = q(r) \label{YE} = \int\limits_{0}^{r} {\rho} _e(x) g(x) x^2 dx \;.
  \end{eqnarray}
 In (\ref{YE}), the second equality says that the electric charge    can also be  obtained from
 the charge density $\rho_e(r)$ of the star. 
 Then the Gravitational field equations (\ref{gfe}) lead to the following differential equations for the metric (\ref{metric}) and electromagnetic field (\ref{electromagnetic1}) of the radiation fluid star $\rho = 3 p$
 \begin{eqnarray}
\frac{1}{\kappa^2 g^2}(\frac{f''}{f}   -\frac{f'g'}{fg}  + \frac{2f'}{rf}  + \frac{2g'}{rg}  + \frac{g^2-1}{r^2}  ) &=&  YE^2   + \rho   \;, \label{gd1}\\
\frac{1}{\kappa^2 g^2}(\frac{f''}{f}   -\frac{f'g'}{fg}  - \frac{2f'}{rf}  - \frac{2g'}{rg}  + \frac{g^2-1}{r^2}  ) &=& YE^2   - \rho/3   \;, \label{gd2}\\
\frac{1}{ \kappa^2 g^2}(   \frac{f''}{f} - \frac{f'g'}{fg} + \frac{g^2 -1 }{r^2}  )   &=& Y E^2   + \rho/3   \;, \label{gd3}
\end{eqnarray}
and we have the following conservation relation from the covariant exterior derivative of the gravitational field equations  (\ref{DtauaJ})
\begin{eqnarray}\label{gd4}
p' + 4 p\frac{f'}{f} = 2(YE)'E + \frac{4YE^2}{r} ,\;
\end{eqnarray}
together with the constraint  from (\ref{YRFk})
 \begin{eqnarray}\label{cond2}
 \frac{dY}{dR}  =  \frac{1}{ 2\kappa^2 E^2}  \;
\end{eqnarray}
where the curvature scalar is 
\begin{eqnarray}\label{Ricci}
R = \frac{2}{g^2} \left( -   \frac{f''}{f}  + \frac{f' g' }{fg} -  \frac{2f'}{fr}  +   \frac{2g'}{gr}  +  \frac{g^2 -1 }{r^2}   \right) \;.
\end{eqnarray}

\subsection{EXACT SOLUTIONS UNDER CONFORMAL SYMMETRY}
We assume that 
 the existence of a one parameter  group of conformal motions for the metric (\ref{metric})
  \begin{eqnarray}\label{conf}
 L_\xi g_{ab} = \phi(r) g_{ab}
 \end{eqnarray}
 where $ L_\xi g_{ab}$ is Lie derivative of the interior metric  with respect to the vector field $\xi$  and $\phi(r)$ is an arbitrary function of $r$. 
 The interior gravitational field  of  stars  can be described  by using this symmetry 
  \cite{mak-harko-2004},\cite{herrera1,herrera2,herrera3}.
  The metric functions $f^2(r) $ and $g^2(r)$    satisfying this symmetry   were obtained 
     as
\begin{eqnarray}\label{s1}
f^2(r) &=& a^2 r^2, \hskip 3 cm 
g^2(r) = \frac{\phi^2_0}{\phi^2}
\end{eqnarray}
  in \cite{herrera1} where  $a$ and $  \phi_0 $ are integration constants. Introducing a new variable $X = \frac{\phi^2}{\phi_0^2}$   in (\ref{s1}) and using this symmetry, equations  (\ref{gd1})-(\ref{cond2})  turn out to be the three differential equations
\begin{eqnarray}
   -\frac{X'}{\kappa^2 r} + \frac{2X +2}{\kappa^2 r^2} - 2\rho &=& 2 YE^2\;, \label{d1}
\\
\frac{X'}{\kappa^2 r} - \frac{2X}{\kappa^2 r^2} +\frac{2\rho}{3} &=& 0  \;,\label{d2}\\
\label{d3}
 p' +  \frac{4p}{r} - 2(YE)'E - \frac{4YE^2}{r} &=&0\;.
\end{eqnarray}
Here we note that  the constraint (\ref{cond2}) is not an independent equation  from (\ref{d1}) and (\ref{d2}), since    we find  the constraint   
  eliminating $\rho$  from (\ref{d1})-(\ref{d2})  and   taking derivative of the result equation  as in \cite{Sert16regular} (where $ \rho=3p$). 
Thus, we have three differential equations (\ref{d1}-\ref{d3}) and four unknowns ($X, \rho, Y, E$). So a given theory or a non-minimal coupling function $Y(R)$, it may be possible to find the corresponding exact solutions for the functions $X, E$ and $\rho$, or inversely, for  a convenient
choice  of any one of the functions   $X, E$ and $\rho$, we may find the corresponding non-minimal theory via the non-minimal function $Y(R)$. In this paper we will continue with the second case offering  physically acceptable  metric solutions.  In the second case,  one of the  challenging problems   is  to solve $r$ from $R(r)$ and re-express the function Y depending on $R$.

 When we choose  the metric function
  $g^2(r) = \frac{1}{X } =\frac{3}{ 1 - br^2}$
    as a result in \cite{mak-harko-2004} with  a constant $b$,  we find the constant curvature scalar $R =4b $  and  a constant non-minimal function $Y(R)$. Then    this model (\ref{action}) turns out that the minimal Einstein-Maxwell case. Therefore we need non-constant curvature scalar to obtain non-trivial  solutions of the non-minimal theory.
    Inspired by \cite{mak-harko-2004},  for  $\alpha> 2$  real numbers and   $b\neq 0$, we  offer  the following metric function 
    \begin{eqnarray}\label{g1}
    g^2(r) = \frac{1}{X} = \frac{3}{ 1 + br^\alpha}
    \end{eqnarray}
    which is regular at origin $r=0$  and giving  the following non-constant  and regular  curvature scalar
\begin{eqnarray} \label{R}
R= - b(\alpha+2) r^{\alpha-2}.
\end{eqnarray}
 We note that if $b=0$  the curvature scalar $R$ becomes zero  and $Y$ is a constant again.  Therefore, here we consider the case with  $b\neq 0$   and  obtain the following   solutions to the equations (\ref{d1}-\ref{g1}) 
  \begin{eqnarray}
 \rho(r) &=& \frac{2- br^\alpha(\alpha-2)}{2\kappa^2 r^2} \label{rho1}
 \;, \\
 Y(r) &=& c \left[  1 +   b(\alpha-2)r^\alpha   \right]^{-\frac{3(\alpha+2)}{2\alpha}} \;, \label{Y1}
 \\
 E^2(r) &=&  \frac{ \left[ 1 + b(\alpha-2)r^\alpha\right]^{ \frac{5\alpha+6}{2\alpha} }} {3c \kappa^2 r^2} \label{E1} \;.
   \end{eqnarray}
 Here  $c $ is a non-zero integration constant and it will be determined by the exterior Einstein-Maxwell Lagrangian (\ref{theoryext})  as $c=1$. Using  the charge-radius relation (\ref{YE}), we calculate the total charge inside the volume  with radius $r$
 \begin{eqnarray}\label{qr1}
 q^2(r) =( YEr^2)^2 =  \frac{c  r^2
 	\left[1+b(\alpha-2)r^\alpha \right]^{-\frac{\alpha+6}{2\alpha} }}{3\kappa^2} \;.
 \end{eqnarray} 
 We see that the charge  is regular at the origin $r=0$  for the theory with $\alpha>2$.
Obtaining by
  the inverse of $R(r)$ from (\ref{R})
  \begin{eqnarray}
  r= (\frac{-R}{\alpha b + 2b} )^{1/(\alpha-2)}
  \end{eqnarray}
  the  non-minimal coupling function is calculated as
    \begin{eqnarray}\label{Y2}
    Y(R)= c \left[ 1 + b(\alpha-2)  (    \frac{-R}{\alpha b +2b}  )^{\frac{\alpha}{\alpha-2}}    \right]^{-\frac{3\alpha+6}{2\alpha}  }  \;.
    \end{eqnarray} 
      The non-minimal function  (\ref{Y2}) turns  to  $Y(R) = c$ 
    for  the vacuum case  $R=0$  and  we can choose  $c=1$ to obtain the known minimal Einstein-Maxwell theory  at the exterior region.
    Thus the Lagrangian of our non-minimal gravitational theory (\ref{action}) 
   \begin{eqnarray}\label{theory}
   L =    
   \frac{1}{2\kappa^2} R*1
   -   \left[ 1+ b(\alpha-2)  (   \frac{-R}{\alpha b +2b}  )^{\frac{\alpha}{\alpha-2} }     \right]^{\frac{-3(\alpha+2)}{2\alpha}} F\w *F  + 2A\wedge J + L_{mat}  + \lambda_a\wedge T^a \ \ \ \ \ \ 
   \end{eqnarray} 
  admits  the following metric  
   \begin{eqnarray}\label{metricin}
   ds^2 = - a^2r^2dt^2 + \frac{3}{1+br^\alpha} dr^2 + r^2(d\theta^2 +sin^2\theta d\phi^2 )
   \end{eqnarray}
   together with the energy density,  electric field  and  electric charge 
    \begin{eqnarray}
   \rho(r) &=& \frac{2- br^\alpha(\alpha-2)}{2\kappa^2 r^2} \label{rho3}
   \;, \\
   E^2(r) &=&  \frac{ \left[ 1 + b(\alpha-2)r^\alpha\right]^{ \frac{5\alpha+6}{2\alpha} }} {3 \kappa^2 r^2} \label{E3} \;,\\
   q^2(r) &=&  \frac{  r^2
   	\left[1+b(\alpha-2)r^\alpha \right]^{-\frac{\alpha+6}{2\alpha} }}{3\kappa^2} \;\label{q2}
   \end{eqnarray}
    under  the conformal symmetry  (\ref{conf}) describing the interior of the radiation fluid star with  $\alpha>2\;$. The parameter $b$ in the model will be determined  from the matching condition  (\ref{prs}) and the parameter $\alpha$ can be determined by the related  observations.

  On the other hand, since the exterior region does not have any matter and  source  the  above  non-minimal Lagrangian (\ref{theory}) 
    turns to
  the following   sourceless  minimal Einstein-Maxwell  Lagrangian, 
   \begin{eqnarray}\label{theoryext}
   L =    
   \frac{1}{2\kappa^2} R*1
   -     F\w *F  + \lambda_a\wedge T^a,
   \end{eqnarray}
    which is the  vacuum case  with $Y(R)=1$  and  the field equations  of the  non-minimal theory (\ref{maxwell1}-\ref{gfe})  turn  to the following  Einstein-Maxwell field equations due to $Y_R=0$
   \begin{eqnarray}
   d* F = 0 \; ,   \ \label{maxwell2}
 \ \hskip 2 cm  \   \   dF= 0 \;, \ 
   \\ \label{gfe2}
   - \frac{1}{2 \kappa^2}  R_{bc}
   \w *e^{abc}   =      \iota^a F \w *F - F \w \iota^a *F 
   \end{eqnarray}
 which  lead to $R=0$ from trace equation and  admit 
  the following Reissner-Nordstrom metric  
   \begin{eqnarray}\label{metricext}
   ds^2 = -(1-\frac{2M}{r} + \frac{\kappa^2Q^2}{r^2})dt^2 + (1-\frac{2M}{r} + \frac{\kappa^2Q^2}{r^2})^{-1}dr^2 + r^2(d\theta^2 +sin^2\theta d\phi^2 ) \; 
   \end{eqnarray}
with the electric field 
   \begin{eqnarray}\label{Eext}
   E(r) = \frac{Q}{ r^2}
   \end{eqnarray} 
   at the exterior region.
   Here $M$ is the total gravitational  mass  and $Q=q(r_b)$ is the total charge of the star.  Since the Ricci scalar is zero for the Reissner-Nordstrom solution, the non-minimal function (\ref{Y2}) becomes  $Y=1$ consistent with the above considerations.    As we see from (\ref{maxwell2}) the excitation 2-form $\mathcal{G}= YF$    is replaced by the Maxwell tensor $F$ at the  exterior vacuum region. 
 In order to see a concrete example of this non-minimally coupled theory  we look on 
   the simplest  case where  $\alpha=3$,  then  the non-minimal   Lagrangian is
   \begin{eqnarray}\label{model3}
   L =     \frac{1}{2\kappa^2} R*1 -   (1 - \frac{R^3}{5^3b^2} )^{-\frac{5}{2} } F\w *F  + 2A\wedge J + L_{mat} + \lambda_a\wedge T^a
   \end{eqnarray}
  and its  corresponding field equations admit    the following interior metric  
   \begin{eqnarray}\label{metricin2}
   ds^2 = - a^2r^2dt^2 + \frac{3}{1+br^3} dr^2 + r^2(d\theta^2 +sin^2\theta d\phi^2 )\;.
   \end{eqnarray} Using  the curvature scalar 
   $ R = - 5b r $, we find the energy density,  electric field and charge   as
 \begin{eqnarray}
 \rho(r) &=&    \frac{1}{\kappa^2 r^2} - \frac{b}{2\kappa^2}r \label{rho2}
 \;,\\
  E^2(r) &=&  \frac{(1 + br^3 )^{ \frac{7}{2}}}{3 \kappa^2 r^2}\; , \label{E2}\\
  q(r) &=& \frac{r^2(1+br^3)^{-3/2}}{3\kappa^2} \;.
  \end{eqnarray}
  For  the  exterior region ($R=0$),     the model (\ref{model3})    turns  to the well known Einstein-Maxwell theory which admits the above Reissner-Nordstrom solution.

\section{Matching Conditions}

We will match the interior and exterior metric   (\ref{metricin}), (\ref{metricext}) 
  at the boundary  of the star $r=r_b$   for continuity of the gravitational potential   
\begin{eqnarray}
 a^2r_b^2 &=& 1-\frac{2M}{r_b} + \frac{\kappa^2 Q^2}{r_b^2}\label{a2} \;, \\
 \frac{3}{1 + br_b^\alpha} &=& (  1-\frac{2M}{r_b} + \frac{\kappa^2 Q^2}{ r_b^2} )^{-1}  \label{b0}  \;.
\end{eqnarray}
The matching  conditions    (\ref{a2}) and  (\ref{b0}) give
\begin{eqnarray}
a^2 = \frac{\kappa^2 Q^2   - 2Mr_b  + rb^2  }{r_b^4}\;,\\
b =\frac{ 2r_b^2  - 6Mr_b  + 3\kappa^2Q^2  }{r_b^{2+\alpha}}\;. \label{b11}
\end{eqnarray}
 The vanishing   pressure condition       at the boundary   $r_b$  requires that
\begin{eqnarray}
 p(r_b)  = \frac{2- b(\alpha-2)r_b^\alpha}{6\kappa^2 r_b^2}    = 0   \;, \label{prs}
\end{eqnarray}
 and it determines the  constant $b$  in the non-minimal model   (\ref{theory})  as 
 \begin{eqnarray}
 b = \frac{2}{(\alpha-2)r_b^\alpha}
 \label{b1}\;.
 \end{eqnarray}
  The interior  region of the star   can be considered as a specific medium and the exterior region as a vacuum. Then 
   the excitation 2-form    $\mathcal{G}=YF  $ in the interior    turns  to the Maxwell tensor $F$ at the exterior,  because of  $Y=1$ in this vacuum  region. That is, we  use the continuity of the tensor   at the boundary which leads  to the continuity of the total charge in which  a volume V. Then the total charge for the exterior region is obtained from the Maxwell equation  (\ref{maxwell2}), $d*F=0$, taking the integral $\frac{1}{4\pi}\int_{\partial V} *F= Er^2= Q$,  while the  total  charge in the interior region is given by (\ref{YE}). Thus
  the total charge  $Q$ is  determined by setting $r=r_b$ in (\ref{q2}) as  a last matching condition
 \begin{eqnarray}
 q^2(r_b)  =  \frac{  r_b^2
	\left[1+b(\alpha-2)r_b^\alpha \right]^{-\frac{\alpha+6}{2\alpha} }}{3\kappa^2} &=& Q^2 \;. \label{Qb}
 \end{eqnarray}
  Substituting   (\ref{b1})  in (\ref{Qb}) we obtain the following   total charge-boundary radius relation   
 \begin{eqnarray}
  Q^2  &=&  \frac{  r_b^2  }{ \kappa^2 3^{\frac{3\alpha+6}{2\alpha} }}   \;. \label{Q1}
 \end{eqnarray} 
The ratio $\frac{\kappa^2Q^2}{r_b^2}=3^{-\frac{3\alpha+6}{2\alpha} }$ which is obtained from (\ref{Q1})   is  plotted in Figure 1  depending on the parameter $\alpha$ of the model for different $\alpha $ intervals. As we see from (\ref{Q1})
the charge-radius ratio has the upper limit
\begin{eqnarray}
\frac{\kappa^2 Q^2}{r_b^2} = \frac{1}{3\sqrt{3}} \approx 0.1924 \;.
\end{eqnarray}

  \begin{figure}[t]{}
	\centering
	\subfloat[ $2<\alpha<10 $ ]{ \includegraphics[width=0.4\textwidth]{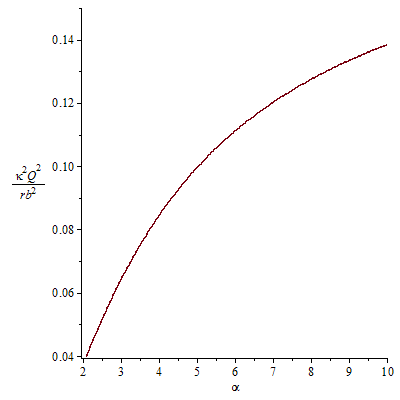} }
	\subfloat[ $2<\alpha<200 $  ]{ \includegraphics[width=0.4\textwidth]{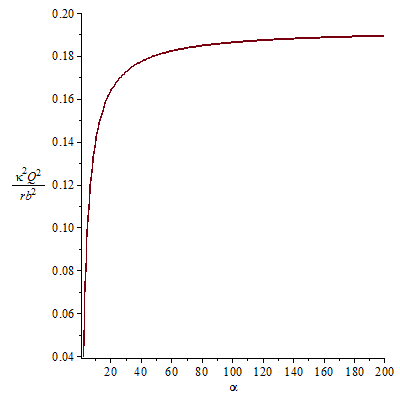}} 
	\\
	\parbox{6in}{\caption{{{\small{  The square of the charge-radius ratio versus the parameter $\alpha$. 
	}}}}}
\end{figure}

 When we compare (\ref{b1}) with (\ref{b11}), we find 
 the following mass-charge relation for the model with the non-minimally coupled electromagnetic fields to gravity
 \begin{eqnarray}
 M= (\frac{\alpha-3}{\alpha-2})\frac{r_b}{3} + \frac{\kappa^2 Q^2}{2r_b} \; \label{M1} \;.
 \end{eqnarray} 
Substituting  the total charge   (\ref{Q1}) in 
  (\ref{M1})  we find the total mass of the star depending on the boundary radius $r_b$ and the parameter $\alpha$ of the model 
 \begin{eqnarray}
M= (\frac{\alpha-3}{\alpha-2})\frac{r_b}{3} + 3^{-\frac{3\alpha+6}{2\alpha} }\; \frac{r_b}{2} \;. \label{M2}
\end{eqnarray}
This mass-radius relation is shown in Figure 2 for  two different $\alpha$ intervals. Taking the limit $\alpha\rightarrow \infty\;$, we can find the upper bound for the mass-radius ratio 
\begin{eqnarray}
\frac{M}{r_b}  < \frac{1}{3} + \frac{1 }{6\sqrt{3}}  \approx  0.4295\; 
\end{eqnarray}
which is slightly smaller than  Buchdahl  bound \cite{Buchdahl2} and  
the bound  given in \cite{Mak1}   for    General Relativistic charged objects.
 \begin{figure}[h]{}
	\centering
	\subfloat[$3<\alpha<10 $ ]{ \includegraphics[width=0.4\textwidth]{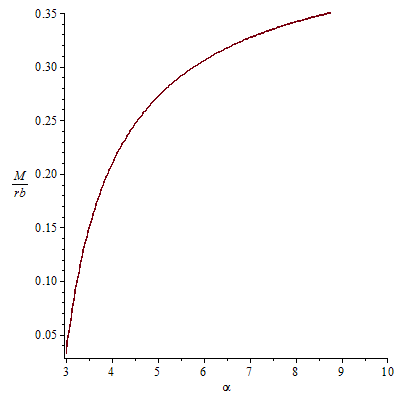} }
	\subfloat[  $3<\alpha<100 $ ]{ \includegraphics[width=0.4\textwidth]{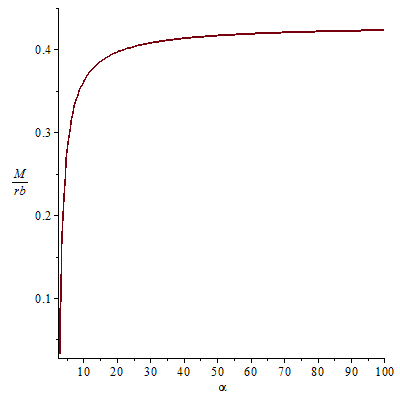}} 
	\\
\parbox{6in}{\caption{{{\small{  The gravitational  mass-radius ratio versus the parameter $\alpha$. 
}}}}}
\end{figure}
 Also, the  matter mass component of the radiation fluid star   is  obtained   from the following integral of the energy density $\rho$
 \begin{eqnarray}\label{Mm1}
 M_m = \frac{\kappa^2}{2} \int_0^{r_b}  \rho r^2 dr =   \frac{\alpha}{2(\alpha+1)} r_b\;.
 \end{eqnarray}
The upper bound of the  matter mass for the radiative star is found as   $      M_m =   \frac{r_b}{2} \; $
taking by the limit $ \alpha \to \infty$.
 The dependence of the matter mass-radius ratio  on the parameter $\alpha $  can be seen in Figure 3.
   \begin{figure}[h]{}
 	\centering
 	\subfloat[  $2<\alpha<100 $ ]{ \includegraphics[width=0.4\textwidth]{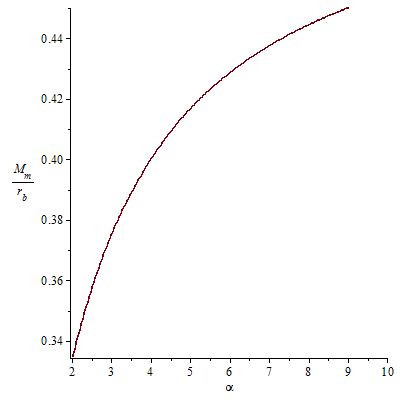} }
 	\subfloat[   $2<\alpha<100 $ ]{ \includegraphics[width=0.4\textwidth]{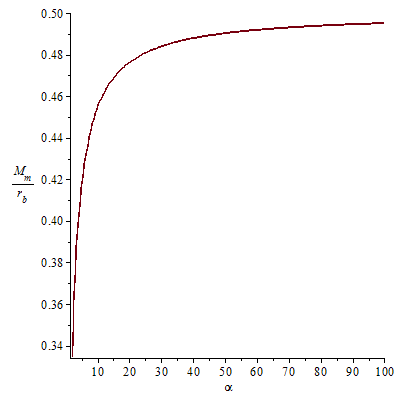}} 
 	\\
\parbox{6in}{\caption{{{\small{  The  matter mass-radius ratio versus the parameter $\alpha$. 
}}}}}
 \end{figure}
Here we emphasize  that each   different  value of $\alpha$ corresponds to a different  non-minimally coupled theory in (\ref{theory}) and the  each different   theory   gives a  different mass-radius relation.\begin{figure}[t]{}
	\centering
	\subfloat[ $3\leq \alpha  < 10 $ ]{ \includegraphics[width=0.4\textwidth]{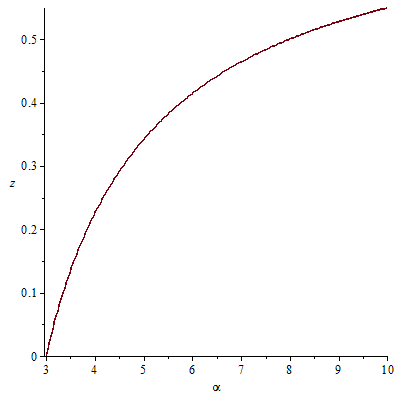} }
	\subfloat[ $3\leq \alpha  < 100 $  ]{ \includegraphics[width=0.4\textwidth]{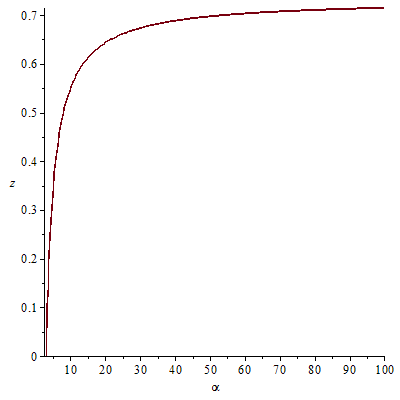}} 
	\\
	\parbox{6in}{\caption{{{\small{  The  gravitational surface redshift  versus the parameter $\alpha$.
	}}}}}
\end{figure}
Additionally, the   gravitational surface redshift $z$  is calculated from
  \begin{eqnarray}\label{z}
  z= (1-\frac{2M}{r_b} + \frac{\kappa^2 Q^2}{r_b^2})^{-
  	\frac{1}{2}} -1  = \sqrt{\frac{3(\alpha-2)}{\alpha} } -1 \;.
  \end{eqnarray}
  
  Taking the limit $\alpha\rightarrow \infty$,
  the inequality  for the redshift is found as
  $z < \sqrt{3} -1 \approx 0.732$ which is smaller than  the bound given in \cite{Mak1} and the Buchdahl bound $z \leq 2$.
We plot the redshift depends on the $\alpha$ in Fig. 4.

For the  case  $\alpha=3$,    
  we calculate all the parameters as $M= \frac{\sqrt{3} r_b}{54}\approx 0.032 r_b   $, \  $ M_m = \frac{3r_b}{8}=0.375r_b$, \   $ Q^2=\frac{\sqrt{3} r_b^2  }{ 27 \kappa^2} \approx \frac{0.064r_b^2 }{\kappa^2}  $\ 
from (\ref{M2}),  (\ref{Mm1}),    (\ref{Q1}).  In this case, because of $\frac{2M}{r_b} = \frac{\kappa^2 Q^2}{r_b^2}$  at the boundary, the metric functions are equal to $1$,  $f(r_b) = g(r_b) = 1 $.
 This means that the total gravitational mass $M$  together with the energy of the electromagnetic field inside the boundary  is exactly balanced by the energy of the electromagnetic field outside the boundary  and then   the  gravitational surface redshift becomes zero from (\ref{z}) for this case $\alpha=3$.
  \begin{table}[t]
 	{\small 
 		\centering
 		\begin{tabular}{|c|c|c|c|c|c|c|}
 			\hline	\ \	 Star  &$\alpha $ & $\frac{M}{r_b}$ & $M (M_\odot)$ & $r_b(km)$ & $\frac{\kappa^2Q^2}{r_b^2}$ &$z$ (redshift) \\
 			\hline
 			SMC X-1  & 3.453 & 0.141 &1.29 & 9.13 & 0.074 & 0.124\\
 			\hline
 			Cen X-3  &  3.555 & 0.157 & 1.49 & 9.51 & 0.076 & 0.145 \\
 			\hline
 			PSR J1903+327 &  3.648 & 0.170 & 1.667 & 9.82 & 0.078 & 0.164\\
 			\hline
 			Vela X-1  & 3.702 & 0.177 &1.77 &9.99 &  0.079& 0.174 \\
 			\hline
 			PSR  J1614-2230 &  3.822 &  0.191 & 1.97 & 10.3 & 0.081 & 0.195 \\
 			\hline
 		\end{tabular}
 		\caption{ Some values of  $\alpha$ and   the corresponding   other  parameters   for some given mass and  mass-radius relations  of    neutron stars}
 		\label{tab:template}
 	}
 \end{table}
In Table 1, we  determine   some $\alpha$  values in the model  for some specific  mass-radius relations in the literature. As we see from  the relation (\ref{M2}) each $\alpha $ gives a mass-radius ratio. Then taking  also the observed  mass values  of  some neutron stars from the literature  we  can find  the corresponding values  of the parameters   such as  the boundary radius,
charge-radius ratio and redshift.

\section{Conclusion}

We have analyzed the exact solutions of the non-minimally coupled $Y(R)F^2$ theory for the  the radiation fluid stars which have the equation of state  $\rho = 3 p$, assuming the existence of a one parameter  group of conformal motions.     We have found new solutions   which lead to regular metric functions and regular Ricci scalar inside the star. We  have obtained  non-negative  matter density $\rho$ and pressure $p$  which vanish at the boundary of the star $r=r_b$, $\rho = 3 p = \frac{r_b^{\alpha} -r^\alpha  }{\kappa^2 r^2 r_b^2}$. The  derivatives of the density and pressure are negative as required for acceptable interior solution,  that is,   $\frac{d\rho}{dr} = 3 \frac{dp}{dr} = -    \frac{(\alpha-2)r^\alpha + 2r_b^\alpha}{\kappa^2r_b^\alpha r^3} $ (where $\alpha >2$).
The speed of sound   $(\frac{dp}{d\rho } )^{1/2} = \frac{1}{\sqrt{3}} < 1$   satisfies the implication of causality, since it does not exceed the speed of light $c=1$. But the mass density $\rho$ and charge density $\rho_e$  have singularity  at the center of the star as the same feature   in \cite{mak-harko-2004}.
However, this feature
is  physically acceptable since the total charge and mass became
finite for the model.

After obtaining the exterior and interior metric solutions of the  non-minimal theory, we  matched them at the boundary $r_b$. Using the vanishing pressure   condition and total charge at the boundary, we obtained  the square of  the total charge-radius ratio $\frac{\kappa^2Q^2}{r_b^2}$,  mass-radius ratio $\frac{M}{r_b}$ and gravitational surface redshift $z$   depending on the parameter $\alpha$ of the model. Taking  the limit $\alpha\rightarrow \infty$, we  found   the ratio $\frac{\kappa^2Q^2}{r_b^2}$ which has the upper bound $\frac{1}{3\sqrt{3} }\approx 0.1924$   and 
the   mass-radius ratio  which has  the upper bound
 $
\frac{M}{r_b}  = \frac{1}{3} + \frac{1 }{6\sqrt{3}}  \approx  0.4295
\;$.
We note that this maximum mass-radius ratio  is smaller than     the  bound  which was  found by Mak {\it   et al.} \cite{Mak1}  for charged  General Relativistic objects    even also Buchdahl bound  $4/9$ \cite{Buchdahl2}  for uncharged  compact objects.
Also  we   found  the upper limit 
$z = \sqrt{3} -1 \approx 0.732$  for the gravitational surface redshift      in  the non-minimal model and it satisfies  the bound given in  \cite{Mak1} for charged stars.
On the other hand  the  minimum redshift $z=0$ corresponds to   the parameter    $\alpha=3$. We have plotted  all these quantities depend on the parameter $\alpha$. 

We   determined    some values of the parameter  $\alpha$   in  Table 1 for some specific  mass-radius relations  which were given by the literature.    Also using the observed mass values we found  the corresponding parameters such as the boundary radius,  
charge-radius ratio and redshift  for some   known  neutron stars.
It would be interesting to generalize the analysis to  the  extended theories of gravity \cite{danila-harko,Odintsov2}  coupled to the  Maxwell theory in future studies.

\section*{Acknowledgement}
I would like to thank to the anonymous referee for
 very useful comments and suggestions.


\end{document}